\documentclass[%
reprint,
amsmath,amssymb,
aps,
]{revtex4-1}
\usepackage{mathrsfs,bm}
\usepackage{subfigure}
\usepackage{slashed}
\usepackage{epstopdf}
\usepackage{graphicx}% Include figure files
\usepackage{dcolumn}% Align table columns on decimal point
\usepackage{bm}% bold math
\begin{document}

\title{Explicit calculation on two-loop corrections to the chiral magnetic effect with NJL model}
\author{Kit-fai Chu, Peng-hui Huang and Hui Liu}
\email{tliuhui@jnu.edu.cn}

\affiliation{Physics Department and Siyuan Laboratory, Jinan University,
	Guangzhou 510632, China}

\date{\today}

\begin{abstract}
Chiral Magnetic Effect(CME) is usually believed not receiving higher order corrections 
		 due to the non-renormalization of AVV triangle diagram in the framework of 
		 quantum field theory. However, the CME-relevant triangle, which is obtained by expanding the current-current correlation requires zero momentum on the axial vertex, is not equivalent to the general AVV triangle when taking the zero-momentum limit owing to the infrared problem on the axial vertex. Therefore, it is still significant to check if there exists perturbative higher order corrections to the current-current correlation. In this paper, we explicitly calculate the two-loop corrections of CME within NJL model with Chern-Simons term which ensures a consistent $\mu_5$. The result shows the two-loop corrections to the CME conductivity are zero, which confirms the non-renomalization of CME conductivity. 
		 
\end{abstract}

\pacs{11.30.Rd,12.38.Mh,11.10Wx}

\keywords{Chiral Magnetic Effect, Quark gluon plasma, Finite temperature field theory}

\maketitle

%гнгнгнгнгнгнгнгнгнгнгнгнгнгнгнгнгнгнгнгнгнгнгнгнгнгнгнгнгнгнгнгнгнгнгнгнгнгнгнгнгнгнгнгнгнгнгнгнгнгнгнгнгнгнгнгнгнгнгнгнгнгнгнгнгнгнгнгнгнгнгнгнгнгнгнгнгнгнгнгнгн

\section{Introduction}
\label{sec:intro}
	 
 The electric current induced by  strong magnetic field and chirality imbalance in heavy ion collisions, 
 which is called chiral magnetic effect(CME)\cite{Kharzeev2014The,Kharzeev2006Parity,Kharzeev2012Charge,Kharzeev2008The,Fukushima2008Chiral,Kharzeev2009Chiral}, is rising interest in recent years. It states 
 that in off-central heavy ion collisions, a strong magnetic field perpendicular  to the collision 
 plane has been generated to induce an electric current due to the non-trivial QCD vacuum configuration \cite{Kharzeev2008The,Mclerran1991Sphalerons} which is 
 described by 
 \begin{equation}
 	n_w=-{{N_fg^2} \over {32\pi^2}} \int d^4x\epsilon_{\mu\nu\rho\lambda}F^l_{\mu\nu}F^l_{\rho\lambda} 
 \end{equation}
 where a non-zero winding number $n_w$ indicates the imbalance of left-handed and right-handed quarks. 
 Since the spin magnetic moment always tends to be parallel to the external magnetic field by the lowest Landau level, 
 the positive(negative) helicity quark carries current parallel(anti-parallel) to its magnetic moment. Hence, the 
 direction of induced current depends on quarks with positive or negative helicity in majority. As a result, an 
 electric current is induced by the separation of quarks carried opposite electrical charge due to the non-zero axial charge 
 density with $P$ and $CP$ violation. The experiments in RHIC\cite{Abelev2009Azimuthal,Collaboration2009Observation,STAR2013Fluctuations,Adamczyk2014Beam} and LHC\cite{Abelev2013Charge} have reported the some observations of charge seperation which might be relevant to CME current.

 The CME classical result, i.e., the linear relationship between the induced current and the magnetic field is often written as 
 \begin{equation}\label{classical}
 	\textbf{J}=\eta{e^2\over2\pi^2}\mu_5 \mathcal{B}
 \end{equation}  
 where $\eta=N_c\sum\limits_f q_f^2$ , $q_f$ is the charge number of flavour $f$ and $\mu_5$ is the axial 
 chemical potential. This result can be achieved in various 
 methods, such as balancing the energy, solving the Dirac equation and from the thermal potential or the 
 effective action\cite{Fukushima2008Chiral}. It is also related to the AVV triangle diagram which contains an axial 
  vertex and two vector vertices. The relation of triangle diagram to CME also analyzed in the 
  longitudinal and transverse part of anomalies\cite{Buividovich2013Anomalous}. Moreover, the CME can also be studied in the holographic 
  model\cite{Yee2009Holographic,Rebhan2010Anomalies,Gynther2011Holographic,Gorsky2011Chiral}, anomalous hydrodynamics\cite{Son2009Hydrodynamics} and lattice simulation\cite{Buividovich2009Numerical,Yamamoto2011Chiral}. 
 
   \begin{figure}[b]
 	\includegraphics[width=0.8\linewidth]{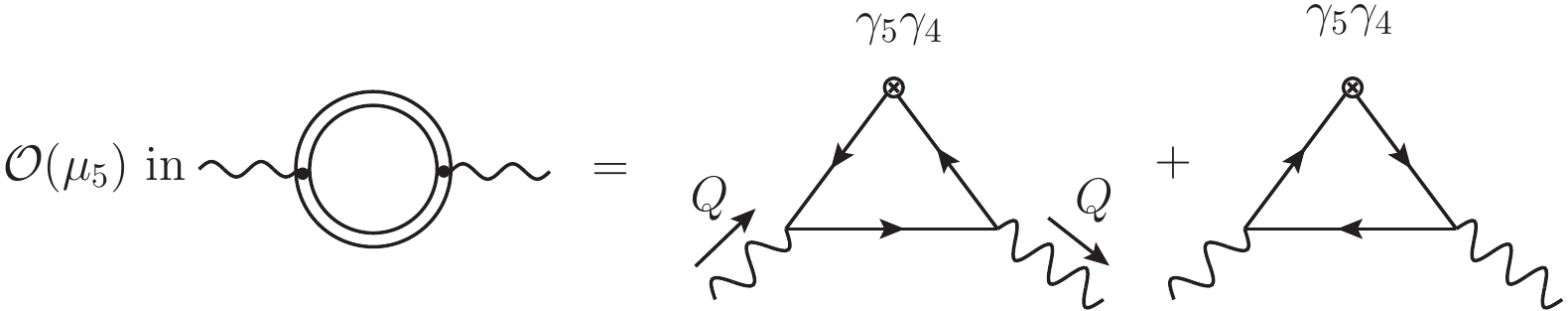}
 	\caption{The sketch of expansion of current-current correlation with respect to the CME. The double line denotes for the propagator with $\mu_5$, while the single line denotes for the regular fermion propagator. The cross vertex denotes for the axial vertex with zero-momentum incoming. }
 	\label{fig:sketch}  	
 \end{figure}
 
 The non-renormalization of CME is a rather subtle issue in current publications.  In the framework of quantum field theory, the induced electric current can be related to the magnetic field through linear response theory. Therefore the CME conductivity is proportional to the current-current correlation which contains the parameter $\mu_5$ that discribes the imbalance of chirality. Expanding the correlation to the first order of $\mu_5$ is equivallent to tranform the two-point loop diagram to the VVA triangle diagram(see fig.\ref{fig:sketch}, and also Sec.\ref{sec:framwork}), which is protected from higher order corrections through the well-known Adler-Bardeen theorem\cite{Adler1996Absence}.  In addition, even introducing Chern-Simons term into the effective action for a consistent $\mu_5$\cite{Rubakov2010On}, which guarantees the conservative axial charge,  one could also prove that all corrections to the topological mass term vanish identically\cite{Coleman1985No}. Such non-renomalization property agrees with the  hydrodynamic calculations\cite{Son2009Hydrodynamics,Isachenkov2011The} that makes people believe the CME recieves no higher order perturbative corrections. However, this is not the whole story. From the lattice point of view,  the lattice simulation for CME disagrees with the classical CME result in quantitative level\cite{Buividovich2009Numerical,Yamamoto2011Chiral}, even though the systematic effects have been considered\cite{Yamamoto2011Lattice}. Kinetic theory points out that those differences may come from the  attractive axial vector interaction\cite{Zhang2012Correction}. The interacting lattice model indicates that CME may receive correction from  inter-fermion interactions which are not relevant in practice\cite{Buividovich2015Chiral}. From the quantum field theory point of view, there is an axial vertex with zero momentum on the triangle diagram with respect to the classical CME, which is not equivallent to the general AVV triangle when taking the zero-momentum limit on the axial vertex because it suffers IR problem. As we know, only at the limit order $\lim\limits_{{\bf q}\rightarrow 0}\lim\limits_{q_0\rightarrow 0}$, with $(q_0,{\bf q})$ the four-momentum of axial vertex, the general AVV triangle can reproduce the classical CME result\cite{Hou2011Some}. What is more, if introducing the Chern-Simons term, one could prove that the current-current correlation with respect to the CME, which is represented by the AVV triangle with zero incoming momentum at axial vertex, vanishes at one-loop level\cite{Hou2011Some}, so that Eq.(\ref{classical}) is completely contributed by the Chern-Simons term. As far as we know, there is no general argument which suggests  the current-current correlation vanishing for all higher order corrections. The full picture of the higher order correction of chiral magnetic effect is ambiguous yet. In this paper, we aim to calculate the current-current correlation which contributes to the CME current at two-loop level within NJL model to examine whether the two-loop corrections exist or not.

 In section II, we will start from the framework of the chiral magnetic conductivity through the 
 thermal field theory. In section III, we will calculate the two-loop diagrams from the NJL model with 
 Pauli-Villars regularization. Section IV is the conclusion. 
 In this paper, we will adopt the Euclidean metric diag(1,1,1,1) and the Minkowski four momentum 
 $P=(\textbf{p},ip_0)$ for $p_0$ real. All gamma matrices are hermitian.

%гнгнгнгнгнгнгнгнгнгнгнгнгнгнгнгнгнгнгнгнгнгнгнгнгнгнгнгнгнгнгнгнгнгнгнгнгнгнгнгнгнгнгнгнгнгнгнгнгнгнгнгнгнгнгнгнгнгнгнгнгнгнгнгнгнгнгнгнгнгнгнгнгнгнгнгнгнгнгнгн

\section{The framework of chiral magnetic conductivity}
\label{sec:framwork}
	  
	  Consider the effective Lagrangian density of a massless quark matter with non-zero  axial charge :		%Section2
	  \begin{widetext}
	  \begin{equation}
	  	\begin{aligned}
	  		\mathcal{L}=-{1\over4}F^l_{\mu\nu} F^l_{\mu\nu}-{1\over4}F_{\mu\nu}F_{\mu\nu} 	
	  		-\bar{\psi} \left( {\gamma_\mu}{\partial_\mu}  -i g T^l \gamma_\mu A^l_\mu -\dot{\imath} e \hat{q} \gamma_\mu A_\mu \right) {\psi}
	  	 +\mu_5(\bar{\psi}\gamma_4\gamma_5\psi +i\Omega_4)
	  	\end{aligned}
	  	\label{eq:eq2.1}
	  \end{equation}
	  \end{widetext}	  
	  where $\hat{q}$ is the diagonal matrix of electric charge in flavour space, $\mu_5$ is the  axial chemical potential respectively. $\Omega_4$ is the fourth component of the 
	  Chern-Simons term which is given by 	  
	  \begin{flalign}
	  	\Omega_\mu=i\frac{N_f g^2}{8\pi^2}\epsilon_{\mu\nu\rho\lambda} A^l_\nu &\left( \frac{\partial A^l_\lambda}{\partial x_\rho}-
	  	\frac{1}{3}f^{lab} A^a_\rho A^b_\lambda\right)\nonumber\\
	  	& + i\eta \frac{e^2}{4\pi^2} \epsilon_{\mu\nu\rho\lambda} 
	  	A_\nu \frac{\partial A_\lambda}{\partial x_\rho}
	  	\label{eq:eq2.2}
	  \end{flalign}
	  where $N_f$ is the number of flavour and $l$ is the colour index for $SU(N_c)$ field $(N_c=3)$. \\

	  In the thermal field theory, the generating functional of  Green's function is corresponding to the partition 					%Brifing
	  function. Following the general procedure of thermal field theory\cite{Hou2011Some}, the electric current can be written as 	  
	  \begin{equation}
	  	J_i(x)={\delta\Gamma [\mathcal{A}]\over\delta \mathcal{A}_i(x)} +\eta{e^2\over2\pi^2}\mu_5 \mathcal{B}_i
	  	\label{eq:eq2.3}
	  \end{equation} 
	  where $\mathcal{A}_i$ and $\mathcal{B}_i$ are the thermal average of gauge field $A_i$ and magnetic field $B_i$. 
	  The second term of Eq.(\ref{eq:eq2.3}) is generated by the Chern-Simons term. Expanding the action 
	  $\Gamma [\mathcal{A}]$  according to  $\mathcal{A}$, one will obtain  the current-current correlation 
	  as leading order coefficient, 	  
	  \begin{equation}
	  	\Gamma [\mathcal{A}]=\int {d^4Q\over(2\pi)^2} \left[-{1\over2} 
	  	\mathcal{A}^*_\mu(Q)\Pi_{\mu \nu}(Q)\mathcal{A}_\nu(Q)+\mathcal{O}(\mathcal{A}^3) \right].
	  	\label{eq:eq2.4}
	  \end{equation}
	  Therefore, the induced current is given by 	  
	  \begin{equation}
	  	J_i (Q)=\mathcal{K}_{ij} \mathcal{A}_j(Q)
	  	\label{eq:eq2.5}
	  \end{equation}  
	  where 	 
	  \begin{equation}
	  	\mathcal{K}_{ij}=-\Pi_{ij}(Q)-i\eta \frac{e^2}{2\pi^2}\mu_5 \epsilon_{ijk}q_k+\mathcal{O}(\mathcal{A}^2).
	  	\label{eq:eq2.6}
	  \end{equation}
	  Regarding to the chiral magnetic conductivity, we need to isolate the coefficient of $\mu_5\epsilon_{ijk}q_k$ in $\Pi_{ij}$. 
	  
\begin{figure}[t]
	  	\centering
		\includegraphics[width=0.7\linewidth]{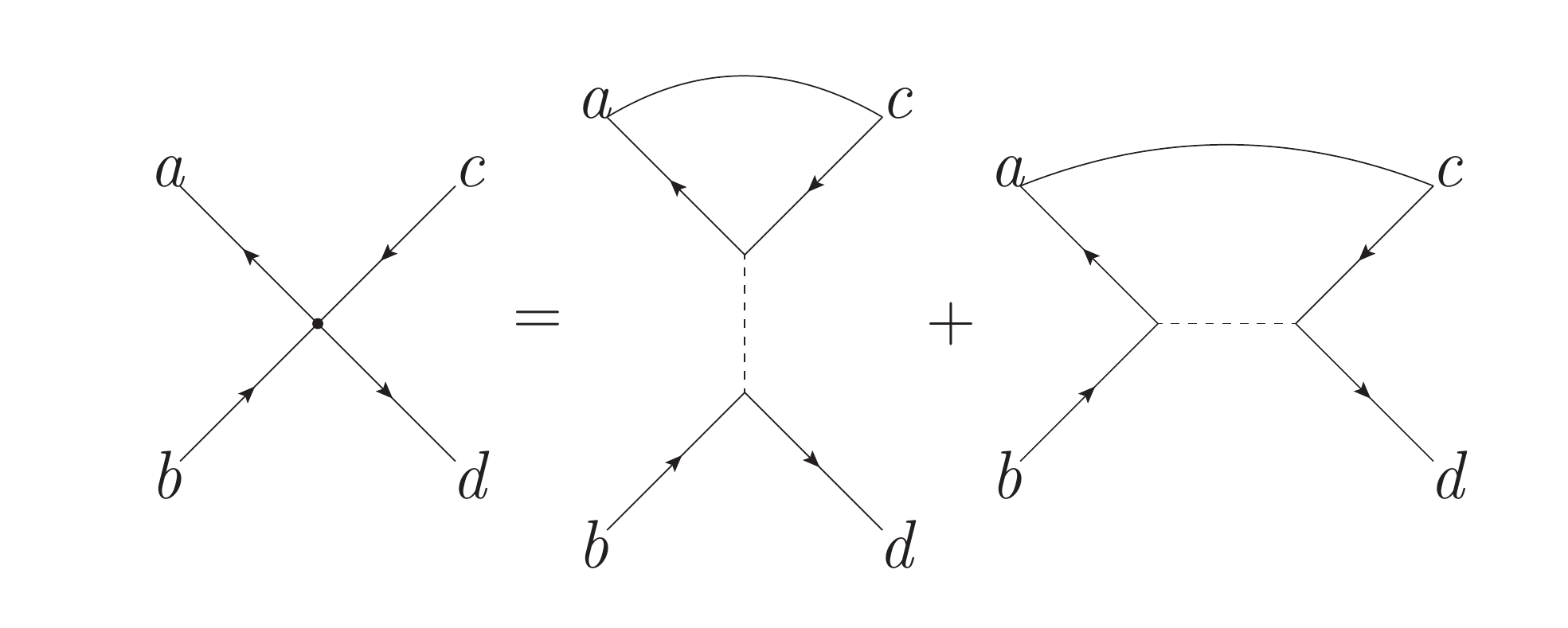}
	  	\caption{The four fermions interaction where the first diagram on the right hand side refers to the direct term and the second diagram refers to the exhange term.}
	  	\label{fig:fig1}  	
	  \end{figure}

\begin{figure}[t]                       																			%diagram              			
	  		\centering
	  		\subfigure[]{\includegraphics[width=0.4\linewidth]{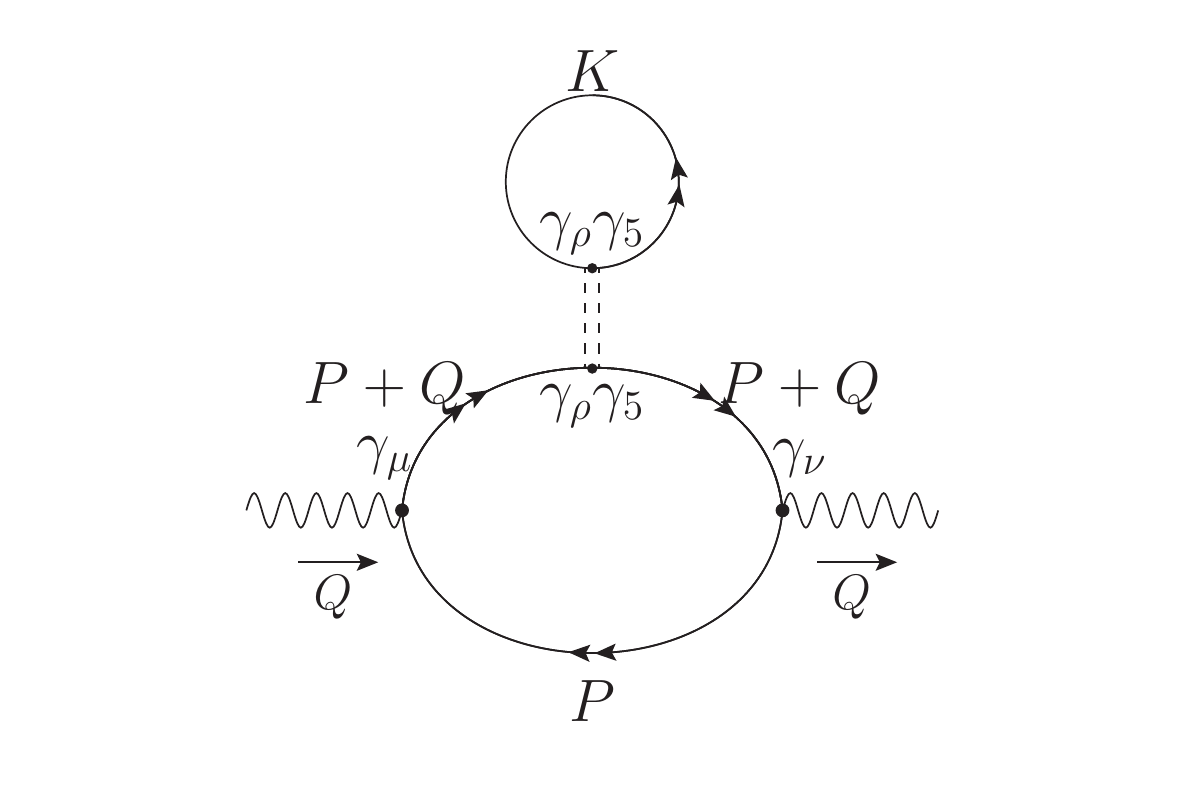}}							 	  								
	  		\subfigure[]{\includegraphics[width=0.4\linewidth]{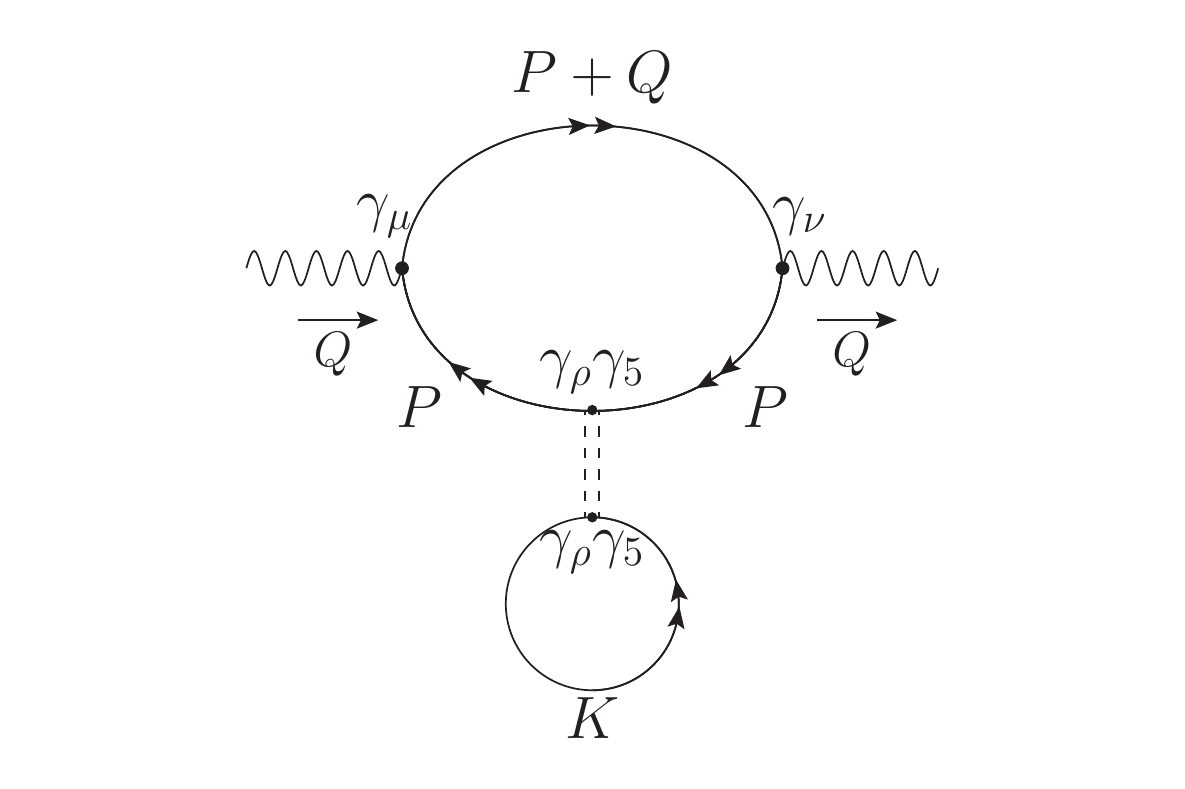}}		
	  		\subfigure[]{\includegraphics[width=0.4\linewidth]{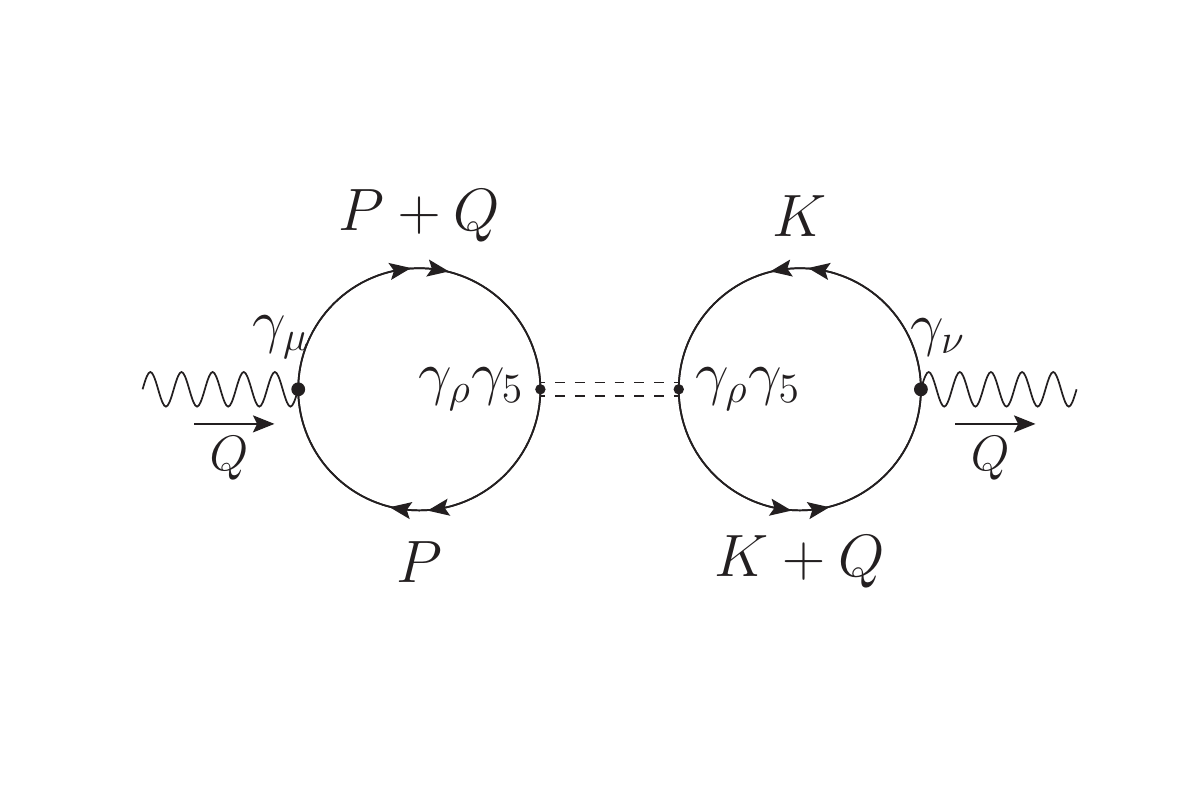}}
	  		\subfigure[]{\includegraphics[width=0.4\linewidth]{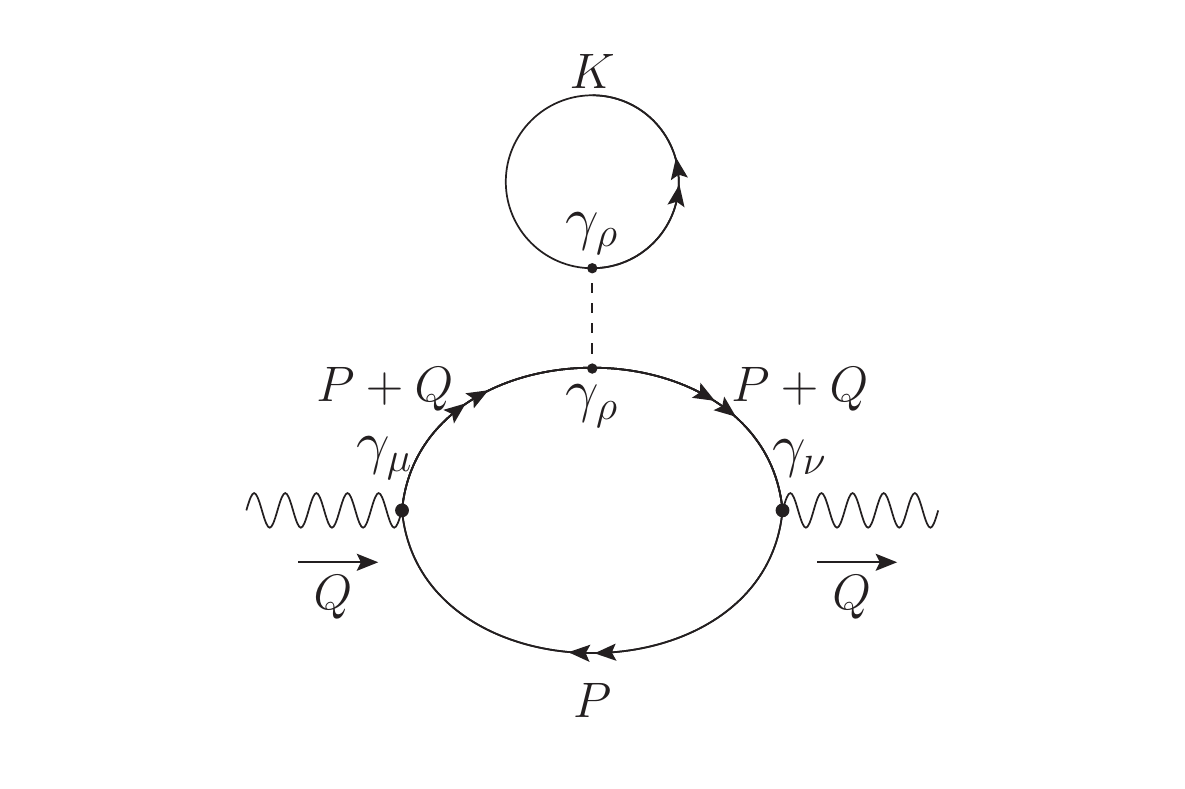}}
	  		\subfigure[]{\includegraphics[width=0.4\linewidth]{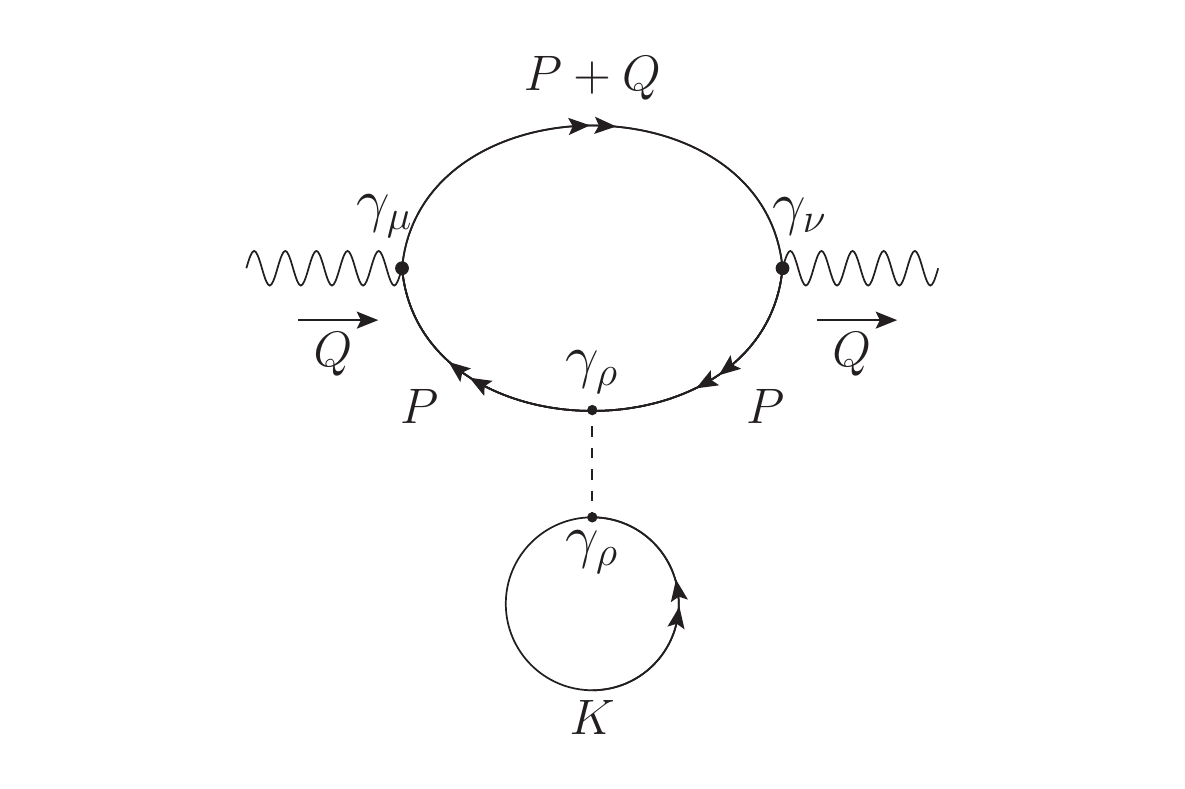}}
	  		\subfigure[]{\includegraphics[width=0.4\linewidth]{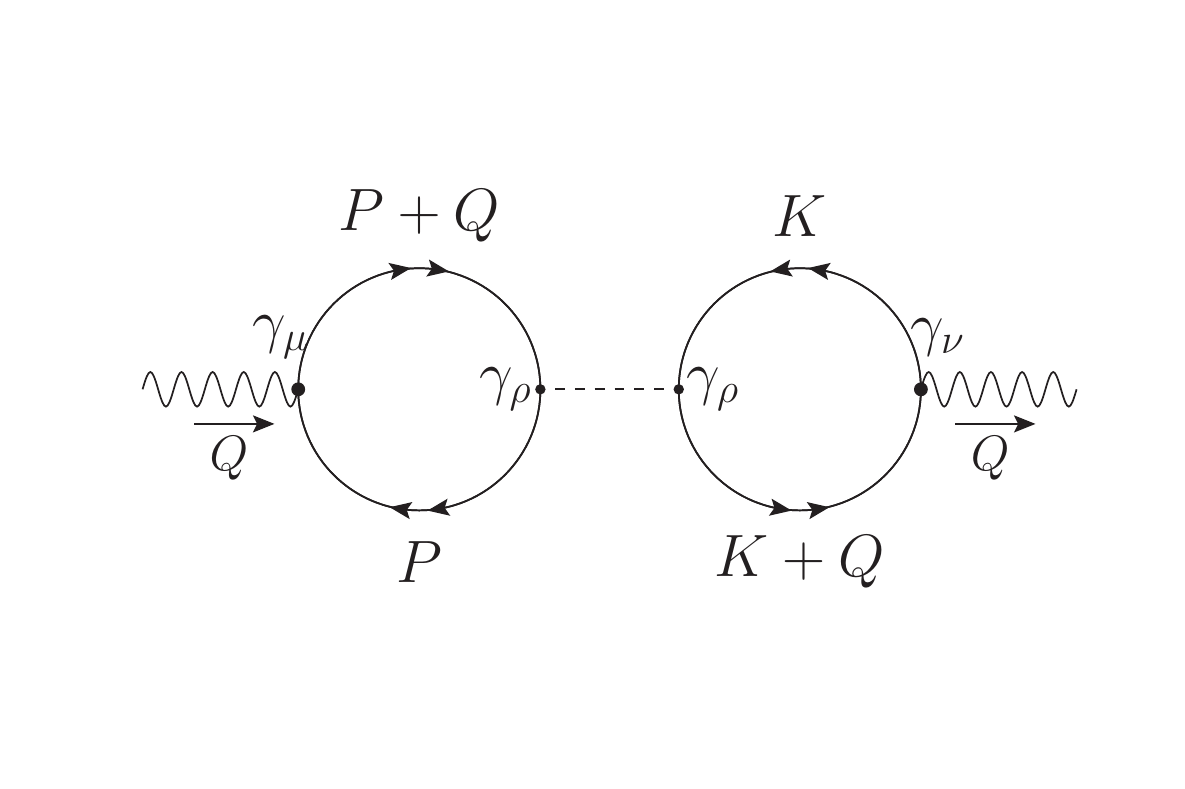}}
	   	\caption{\small{Feynman diagrams of the two-loop current-current correlation. The solid fermion  line 
	  			represents the propagator with $\mu_5$. The dashed line and double dashed line represent the  
	  			coupling of vector vertex and axial vector vertex, respectively.} }
	  	\label{fig:fig2}
	  \end{figure}

	  Obviously, the second term of Eq.(\ref{eq:eq2.6}) is originated from the Chern-Simons term which is protected from higher order corrections. While the first term is a two-point correlation function which may recieve decorations from quantum chromodynamics(QCD). These decorations run rather complicated at two-loop or higher levels, thus we introduce the Nambu-Jona-Lasinio(NJL) model to simulate the QCD interactions where the four-fermion interactions instead of non-Abelian gauge field will greatly reduce the complications in calculation. The interacting part of NJL Lagrangian is given by 	  
	  \begin{equation}
	  	\mathcal{L}^{NJL}_{int}=-GV_\mu(x)V_\mu(x)=-G(\bar{\psi}\gamma_\mu\psi)^2
	  	\label{eq:eq2.7}
	  \end{equation} 
	  with $G$ the coupling constant. Notice that a momentum space cutoff $\Lambda$ is provided in $g^2=G\Lambda^2$. 
	  Notice that the interaction in Eq.(\ref{eq:eq2.7}) contains both direct and exchange terms which corresponds to two types of 
	  contraction which are shown in figure \ref{fig:fig1}. 
	  By introducing the Fierz transformation, one obtains the Lagrangian  
	  \begin{equation}
	  	\mathcal{L}^{NJL}_{int}=G[(\bar{\psi}\psi)^2-(\bar{\psi}\gamma_5\psi)^2]-{3G\over 2}(\bar{\psi}\gamma_\mu\psi)^2 
						  	-{G\over 2}(\bar{\psi}\gamma_\mu\gamma_5\psi)^2
	  	\label{eq:eq2.8}
	  \end{equation}
	  where only direct interactions are involved.  It is easy to verify that only the last two terms of Eq.(\ref{eq:eq2.8}) have non-zero contribution owing to the traces of gamma matrix.  Therefore, all we need to compute are the 6 diagrams in figure \ref{fig:fig2} in which the single or double dashed lines  corresponds to the vector and axial vector direct coupling, not propagators.   

%гнгнгнгнгнгнгнгнгнгнгнгнгнгнгнгнгнгнгнгнгнгнгнгнгнгнгнгнгнгнгнгнгнгнгнгнгнгнгнгнгнгнгнгнгнгнгнгнгнгнгнгнгнгнгнгнгнгнгнгнгнгнгнгнгнгнгнгнгнгнгнгнгнгнгнгнгнгнгнгнгн
	
	\section{The two-loop corrections}
	\label{two loops}
				
		Following the effective Lagrangian Eq.(\ref{eq:eq2.1}), one can read out the free quark 
		propagator with a four momentum $P=(\textbf{p},p_4)=(\textbf{p},ip_0)$ 
		
		\begin{equation}
			S_F(P|m)={i\over{\slashed{P} +\mu_5\gamma_4\gamma_5-m}}
			\label{eq:eq3.1}
		\end{equation} 
		where $m$ is the quark mass and $\slashed{P}=\gamma_4 p_0-i\gamma\cdot\textbf{p}$. In our calculation, we consider light quarks, i.e., the quark mass will be set to zero. But since we involve Pauli-Villars regularization to guarantee the charge conservation, we keep the mass in	the format of propagator. In the following calculation and statement on figure \ref{fig:fig2}, the colour-flavour factor $\eta$ is suppressed and the main figure number is omitted so that figure (a) refers to figure \ref{fig:fig2}(a) so on and so forth. In order to compare with the classical CME conductivity, we  focous on the  static limit  $\omega=0$  and  concern on the term that contains the structure like $\mu_5 q_k\epsilon_{ijk}$ in $\Pi_{ij}$. In the following calculations, the terms that irrelevant with such structure are neglected.

%гнгнгнгнгнгнгнгнгнгнгнгнгнгнгнгнгнгнгнгнгнгнгнгнгнгнгнгнгнгнгнгнгнгнгнгнгнгнгнгнгнгнгнгнгнгнгнгнгнгнгнгнгнгнгнгнгнгнгнгнгнгнгнгнгнгнгнгнгнгнгнгнгнгнгнгнгнгн	

\subsection{Figures (a) and (b)}\label{sec:ab}
		
		Let us begin with  figure (a) and figure (b). Notice that the small loops  with momentum $K$ in these two diagrams are the same, one can extract it out and denote it by $ \Lambda^A_{\rho}$, where the subscript $A$ means axial vertex coupling, and  explain figures (a) and 	(b) as
		\begin{equation}
			\Pi^{a+b}_{\mu\nu} \equiv \Lambda^A_{\rho}  \times \left( \Xi^{A,a}_{\rho\mu\nu} + \Xi^{A,b}_{\rho\mu\nu} \right)
			\label{eq:eq3.2}
		\end{equation}			
		where $\Xi^{A,a}_{\rho\mu\nu}$ and $\Xi^{A,b}_{\rho\mu\nu}$ represent the big loop in figures (a) and (b) respectively. The explicit expression for 
		$ \Lambda^A_{\rho}$, $\Xi^{A,a}_{\rho\mu\nu}$ and $\Xi^{A,b}_{\rho\mu\nu}$ are
		\begin{flalign}
			\Lambda^A_\rho= \frac{G}{2}i T\sum\limits_{k_0}&\int{d^3\textbf{k}\over(2\pi)^3} 														%dia A and B
						 \textnormal{tr}\left[ \frac{i}{\slashed{K}+\mu_5\gamma_4\gamma_5}\gamma_\rho\gamma_5\right. \nonumber\\
						 &+\left. \sum\limits^{\infty}_{s'=1}C_{s'} \frac{i}{\slashed{K}+\mu_5\gamma_4\gamma_5-M_{s'}}\gamma_\rho\gamma_5  \right]
			\label{eq:eq3.3}
		\end{flalign}						 
		and	
		\begin{widetext}	
		\begin{align}
			\begin{aligned}				 
			\Xi^{A,a}_{\rho\mu\nu}(Q)=&i e^2 T\sum\limits_{p_0}\int{d^3\textbf{p}\over(2\pi)^3} 
									\textnormal{tr} \left[  \frac{i}{\slashed{P'}+\mu_5\gamma_4\gamma_5}\gamma_\rho\gamma_5 
									\frac{i}{\slashed{P'}+\mu_5\gamma_4\gamma_5}\gamma_\mu 
									\frac{i}{\slashed{P}+\mu_5\gamma_4\gamma_5}\gamma_\nu \right.\\
									&+\left. \sum\limits^{\infty}_{s=1}C_{s}  
									\frac{i}{\slashed{P'}+\mu_5\gamma_4\gamma_5-M_s}\gamma_\rho\gamma_5 
									\frac{i}{\slashed{P'}+\mu_5\gamma_4\gamma_5-M_s}\gamma_\mu 
									\frac{i}{\slashed{P}+\mu_5\gamma_4\gamma_5-M_s}\gamma_\nu \right]
			\label{eq:eq3.4}														
			\end{aligned}\\
			\begin{aligned}
			\Xi^{A,b}_{\rho\mu\nu}(Q)=&i e^2 T\sum\limits_{p_0}\int{d^3\textbf{p}\over(2\pi)^3} 
								\textnormal{tr}\left[ 
								 \frac{i}{\slashed{P'}+\mu_5\gamma_4\gamma_5}\gamma_\mu 
								\frac{i}{\slashed{P}+\mu_5\gamma_4\gamma_5}\gamma_\rho\gamma_5 
								\frac{i}{\slashed{P}+\mu_5\gamma_4\gamma_5}\gamma_\nu \right.\\
								&+\left. \sum\limits^{\infty}_{s=1}C_{s}  
								\frac{i}{\slashed{P'}+\mu_5\gamma_4\gamma_5-M_s}\gamma_\mu 
								\frac{i}{\slashed{P}+\mu_5\gamma_4\gamma_5-M_s}\gamma_\rho\gamma_5 
								\frac{i}{\slashed{P}+\mu_5\gamma_4\gamma_5-M_s}\gamma_\nu \right]
			\label{eq:eq3.4.2}					
			\end{aligned}			
		\end{align}
		\end{widetext}
		where $P'=P+Q$. Pauli-Villars regulators are involved whose coefficients are restricted by the  condition
		\begin{equation}
			\sum\limits^{\infty}_{s=1}C_s=-1 \qquad \text{or} \qquad \sum\limits^{\infty}_{s=0}C_s=0
			\label{eq:eq3.5}
		\end{equation}  	
		Firstly, we expand the spacial component of Eq.(\ref{eq:eq3.3}) to the linear order of $\mu_5$, and complete the trace and obtain		

		\begin{flalign}
		 	\Lambda^A_i=4GT\mu_5 \sum\limits^{\infty}_{s'=0}C_{s'}\sum\limits_{k_0} \int \frac{d^3\textbf{k}}{(2\pi)^3} 
		 	&\frac{k_4 k_i}{(-k^2_4-\textbf{k}^2-M^2_{s'})^2}\nonumber\\
		 	&\qquad+\mathcal{O}(\mu_5^2)
		 	\label{eq:eq3.6}			
		\end{flalign}	
		where we applied a more compact resummation form with $M_0=0$ and $C_0=1$.					
		 Notice that the sum over Matsubara frequencies is done by the contour integration as shown in Appendix A, one can obtain that 		
		\begin{equation}
			\begin{aligned}
		 	T\sum\limits_{k_0} \frac{k_0 k_i}{(k^2_0-\textbf{k}^2-M^2_{s'})^2}=&\frac{-k_i}{4z^2}f'(-z)+\frac{k_i}{4z^2}f'(z)
			\end{aligned}
			\label{eq:eq3.7}				
		\end{equation}	
		where $f'(z)\equiv \frac{\text{d} f}{\text{d} z}$, denoting for the derivative of the distribution function 
		\begin{equation}
			f(z)=\frac{1}{e^{\beta z}+1} 
		\end{equation}
		and $z=\sqrt{p^2+M^2_s}$. Since $f'(-z)=f'(z)$, the linear $\mu_5$ term of $\Lambda^A_i$ becomes zero.		
		
		Then, we consider the temporal component $\Lambda^A_4$, which is contracted with $\Xi_{4\mu\nu}$. Since the leading order of $\Lambda^A_\rho$ is linear to $\mu_5$,  it is sufficient 
		to consider the zeroth order of $\mu_5$ in $\Xi_{4\mu\nu}$. In the following calculation, we will suppress the index 4 for convenience. The contribution of figure (a) is given by 		

		\begin{flalign}
		 	\Xi_{\mu\nu}^{A,a}(Q)=&ie^2\sum\limits^\infty_{s=0} C_s \times T\sum\limits_{p_0}\int{d^3\textbf{p}\over(2\pi)^3} 	\nonumber\\					  
			 &	\textnormal{tr}\left[ \frac{i}{\slashed{P'}-M_s}\gamma_\rho\gamma_5 \frac{i}{\slashed{P'}-M_s}\gamma_\mu 
							\frac{i}{\slashed{P}-M_s}\gamma_\nu \right].
			\label{eq:eq3.8}				
		\end{flalign}
		The introducing of  series of Pauli-Villars regulator cancels out all UV divergences and there has applied a more compact resummation form with $M_0=0$. After straightforward evaluation on the trace, we expand the expression in terms of $q$ and single out its linear  terms which yields		
		\begin{flalign}
		 	\Xi^{A,a}_{ij}(0,q) =& -4ie^2\epsilon_{ijk}q_k\sum\limits^\infty_{s=0} C_s \times T\sum\limits_{p_0}\int{d^3\textbf{p}\over(2\pi)^3}\nonumber\\	
						 &	\frac{2p^2_4-\frac{2}{3}p^2+2M^2_s}{[-p^2_4-(p^2+M^2_s)]^3}
			\label{eq:eq3.9}			
		\end{flalign}	
		where $\mu=i$, $\nu=j$ and $\int d^3\textbf{p} p_lp_k\rightarrow\int d^3\textbf{p}\frac{1}{3}p^2\delta_{lk}$ are applied.\\
	
		Following the same steps, we can handle with $\Xi^b_{ij}$ and  finally obtain  		
		\begin{flalign}
			\Xi^{A,a}_{ij}+\Xi^{A,b}_{ij} =& 4ie^2\epsilon_{ijk}q_k \sum\limits^\infty_{s=0} C_s \times T\sum\limits_{p_0} \int\frac{d^3\textbf{p}}{(2\pi)^3}\nonumber\\
								&	\frac{-3p^2_4+p^2-3M^2_s}{[-p^2_4-(p^2+M^2_s)]^3}.
			\label{eq:eq3.12}			
		\end{flalign}		
		 After performing the summation on Matsubara frequencies, we obtain that 		
		\begin{flalign}
			\Xi^{A,a}_{ij}+\Xi^{A,b}_{ij} =& 4ie^2\epsilon_{ijk}q_k \int\frac{d^3\textbf{p}}{(2\pi)^3}
								 \left\{\frac{1}{2p} \frac{\beta^2 e^{\beta p}(e^{\beta p}-1)}{(e^{\beta p}+1)^3}\right. \nonumber\\
								 &+\left. \sum\limits^\infty_{s=1} C_s \frac{3M^2_s}{4} \frac{1}{(p^2+M^2_s)^{5/2}} \right\}.		 
			\label{eq:eq3.13}					
		\end{flalign}	
		After performing the three-momentum integration, we obtain	
		\begin{align}
			\int d^3\textbf{p} \frac{1}{2p} \frac{\beta^2 e^{\beta p}(e^{\beta p}-1)}{(e^{\beta p}+1)^3}=\pi\\
			\int d^3\textbf{p} \frac{1}{(p^2+M^2_s)^{5/2}}=\frac{4\pi}{3M^2_s}.
			\label{eq:eq3.13.2}
		\end{align} 
		 Then the first term in Eq.(\ref{eq:eq3.13.2}) cancel each other by considering  Eq.(\ref{eq:eq3.5}), which yields
		\begin{equation}
				\Xi^{A,a}_{ij}+\Xi^{A,b}_{ij} =  0.
				\label{eq:eq3.13.3}	
		\end{equation}		
		Therefore, we end up with
		\begin{equation}
		\Pi^{a+b}_{ij}=0.
		\label{a+b}	
		\end{equation}

	\subsection{Figure (c)}																					%dia C
		Figure (c) contains two similar loops in which each loop is denoted by $\Theta^{A}$. We write it as  	
		\begin{equation}
			\Pi^c_{\mu\nu} \equiv  -\frac{G}{2} \Theta^{A}_{\mu\rho} \times \Theta^{A}_{\rho\nu}
			\label{eq:eq3.24}
		\end{equation} 
		where 		
	
			\begin{flalign}\label{eq:eq3.25}
			&	\Theta^{A}_{\mu\rho}=-eT \sum\limits_{s=0}^\infty C_s\sum\limits_{p_0}\int{d^3\textbf{p}\over(2\pi)^3} \nonumber\\
			&	\textnormal{tr} \left(   \frac{i}{\slashed{P'}+\mu_5\gamma_4\gamma_5-M_s}\gamma_\mu 
				\frac{i}{\slashed{P}+\mu_5\gamma_4\gamma_5-M_s}\gamma_\rho\gamma_5 \right).
			\end{flalign}

	Since we are aiming at the spacial components of current-current correlation, we set  $\mu=i$ and expand it in terms of $\mu_5$ to the linear order as
	
		\begin{equation}
			\begin{aligned}
				&\Theta^{A}_{i\rho}(0,q) =-eT \sum\limits_{s=0}^\infty C_s\sum\limits_{p_0}\int{d^3\textbf{p}\over(2\pi)^3}\left\{ \frac{-4p_4 \epsilon_{i\rho k}q_k}{[-p^2_4-(p^2+M^2_s)]^2}\right. \nonumber\\
						&\left.+ \frac{4i(\frac{8}{3}p^2p_4-2P^2p_4-2M^2_s p_4) \delta_{i\rho}}{[-p^2_4-(p^2+M^2_s)]^3}\mu_5\right\}+\mathcal{O}(\mu_5^2).
		\end{aligned}
		\label{eq:eq3.27}			
		\end{equation}
		Now let us look at the summation of Matsubara frequencies in the zeroth order of $\mu_5$  which reads
		\begin{equation}
			\begin{aligned}
			T\sum\limits_{p_0}\frac{p_0}{[p^2_0-(p^2+M^2_s)]^2}	
					&=\frac{-1}{4z} f'(-z) + \frac{1}{4z} f'(z)
			\end{aligned}
			\label{eq:eq3.28}		
		\end{equation}
		where $f'(z)\equiv \frac{\text{d} f}{\text{d} z}$, denoting for the derivative of the distribution function
		\begin{equation}
			f(z)=\frac{1}{e^{\beta z}+1}
		\end{equation}		
		 and $z=\sqrt{p^2+M^2_s}$. Notice that $f'(-z)=f'(z)$, one can conclude that the two terms in (\ref{eq:eq3.28}) are cancelled out which leads the  linear order of $\mu_5$ of Eq.(\ref{eq:eq3.27}) to be zero.
		
		Notice the two loops of figure (c)  have the same structure thus its linear order of $\mu_5$ vanishes, i.e.,  
		\begin{equation}
			\Pi^{c}_{ij}= -\frac{G}{2} \Theta^A_{i\rho}\times\Theta^A_{\rho j}\sim \mathcal{O}(\mu_5^2).
			\label{c}	
		\end{equation}
		Therefore, we end up with
		\begin{equation}
			\Pi^c_{ij}=0.
		\end{equation}

	\subsection{Figures (d) and (e)}																%dia D and E
		Like what we did in section \ref{sec:ab}, we extract the small loops in figures (d) and (e) and 
		denote it by $\Lambda^V_\rho$, where V means the vector vertex coupling, and explain figures (d) and (e) as 		
		\begin{equation}
			\Pi^{d+e}_{\mu\nu} \equiv \Lambda^V_\rho \times \left( \Xi^{V,d}_{\rho\mu\nu} + \Xi^{V,e}_{\rho\mu\nu}	\right)
			\label{eq:eq3.14}
		\end{equation} 
		where $\Xi^{V,d}_{\rho\mu\nu}$ and $\Xi^{V,e}_{\rho\mu\nu}$ represent the big loop in figures (d) and (e) respectively. 
		The explicit expression for $\Lambda^V_{\rho}$, $\Xi^{V,d}_{\rho\mu\nu}$ and $\Xi^{V,e}_{\rho\mu\nu}$ are		
			\begin{widetext}
		\begin{equation}
			\Lambda^V_\rho= \frac{3G}{2}i T\sum\limits^{\infty}_{s'=0}C_{s'}\sum\limits_{k_0}\int{d^3\textbf{k}\over(2\pi)^3} 																
						\textnormal{tr}\left[ \frac{i}{\slashed{K}+\mu_5\gamma_4\gamma_5-M_{s'}}\gamma_\rho  \right]
			\label{eq:eq3.15}			
		\end{equation}		
		and

			\begin{flalign}
			\Xi^{V,d}_{\rho\mu\nu}(Q)= ie^2 T \sum\limits^{\infty}_{s=0}C_{s}\sum\limits_{p_0}\int{d^3\textbf{p}\over(2\pi)^3} 
							\left( 
									\frac{i}{\slashed{P'}+\mu_5\gamma_4\gamma_5-M_s}\gamma_\rho 
									\frac{i}{\slashed{P'}+\mu_5\gamma_4\gamma_5-M_s}\gamma_\mu 
									\frac{i}{\slashed{P}+\mu_5\gamma_4\gamma_5-M_s}\gamma_\nu \right)
			\label{eq:eq3.16}
			\end{flalign}
	
			\begin{flalign}
			\Xi^{V,e}_{\rho\mu\nu}(Q)= ie^2 T\sum\limits^{\infty}_{s=0}C_{s}\sum\limits_{p_0}\int{d^3\textbf{p}\over(2\pi)^3} 
									\textnormal{tr}
									\left(  
									\frac{i}{\slashed{P'}+\mu_5\gamma_4\gamma_5-M_s}\gamma_\mu 
									\frac{i}{\slashed{P}+\mu_5\gamma_4\gamma_5-M_s}\gamma_\rho
									\frac{i}{\slashed{P}+\mu_5\gamma_4\gamma_5-M_s}\gamma_\nu \right).
			\label{eq:eq3.16.2}										
			\end{flalign}					
		\end{widetext}
	It is easy to check that  the term of linear $\mu_5$ vanishes after the trace, and only zeroth order of $\mu_5$ survived in  Eq.(\ref{eq:eq3.15}).
		However, even in the zeroth order, the spacial components of $\Lambda_\rho^V$ are zero due to the integration on an odd function, thus the only non-zero component is  	
		\begin{equation}
		 	\Lambda^V_4=\frac{3}{2}iGT \sum\limits^\infty_{s'=0} C_{s'}   \sum\limits_{k_0} \int \frac{d^3\textbf{k}}{(2\pi)^3} 
							\frac{4k_4}{-k^2_4-\textbf{k}^2-M^2_{s'}} + \mathcal{O}(\mu_5^2)
			\label{eq:eq3.17}			
		\end{equation}		
	 After accomplishing the summation on Matsubara frequencies, one ends up with
		\begin{equation}
			\begin{aligned}
		  	\Lambda^V_4=-3G \sum\limits^\infty_{s'=0} C_{s'}\int \frac{d^3\textbf{k}}{(2\pi)^3}
		  	\end{aligned} 
		  	\label{eq:eq3.18}			
		\end{equation}		
		 Considering the regularization condition Eq.(\ref{eq:eq3.5}), one can conclude that 
		\begin{equation}
			\Pi_{ij}^{d+e}=0
			\label{d+e}
		\end{equation}
		even without doing the tedious calculation on the big loop of $\Xi^{V}_{\rho\mu\nu}$.

	\subsection{Figure(f)}												%dia F
		
		Figure (f) contains two similar loops in which each loop is denoted by $\Theta^{V}$. Then the diagram is interpreted as 		
		\begin{equation}
			\Pi^f_{\mu\nu} \equiv  -\frac{3G}{2} \Theta^{V}_{\mu\rho} \times \Theta^{V}_{\rho\nu}	
			\label{eq:eq3.19}
		\end{equation}	
		where		
		\begin{flalign}
					& \Theta^{V}_{\mu\rho}(Q)=-e T\sum\limits^\infty_{s=0} C_s\sum\limits_{p_0}\int{d^3\textbf{p}\over(2\pi)^3} \nonumber\\
						& \textnormal{tr} \left[  \frac{i}{\slashed{P'}+\mu_5\gamma_4\gamma_5-M_s}\gamma_\mu 
						 \frac{i}{\slashed{P}+\mu_5\gamma_4\gamma_5-M_s}\gamma_\rho \right].
			\label{eq:eq3.20}
		\end{flalign}
			Expand Eq.(\ref{eq:eq3.20}) with respect to $\mu_5$, one finds		
		\begin{flalign}
			\Theta^{V}_{i\rho} =& -eT\sum\limits^\infty_{s=0} C_s	\sum\limits_{p_0}\int{d^3\textbf{p}\over(2\pi)^3}
							\left\{ \frac{4(\frac{2}{3}p^2-P^2-M^2_s) \delta_{i\rho}}{[-p^2_4-(p^2+M^2_s)]^2} \right.\nonumber\\
							&\qquad \left. + \frac{4i(-3p^2_4+p^2-3M^2_s)\epsilon_{i\rho k}q_k}{[-p^2_4-(p^2+M^2_s)]^3} \mu_5\right\}+\mathcal{O}(\mu_5^2)
			\label{eq:eq3.22} 			
		\end{flalign}
		where we set $\mu=i$.  Notice that the 
		integrand of second term of Eq.(\ref{eq:eq3.22}) is zero which has been proved in the Eq.(\ref{eq:eq3.12}). Since the linear order of $\mu_5$ vanished in one of the two loops, the product of two similar loop does not contain the linear $\mu_5$ and thus has zero contribution to the CME conductivity , namely, 		
		\begin{equation}
			\Pi^{f}_{ij}=0.
			\label{f}
		\end{equation}

		Combine Eq.(\ref{a+b}), (\ref{c}), (\ref{d+e}) and (\ref{f}), we find that 
		\begin{equation}
			\Pi^{a+b+c+d+e+f}_{ij}(0,q)=0
			\label{eq:eq3.30}
		\end{equation}
	which means the contribution from two-loop corrections of current-current correlation is zero so that the classical CME coefficient is completely determined by the Chern-Simons term.	
																																
%гнгнгнгнгнгнгнгнгнгнгнгнгнгнгнгнгнгнгнгнгнгнгнгнгнгнгнгнгнгнгнгнгнгнгнгнгнгнгнгнгнгнгнгнгнгнгнгнгнгнгнгнгнгнгнгнгнгнгнгнгнгнгнгнгнгнгнгнгн	

	\section{Discussion}																			%Discussion
	\label{discussion}							
																												
		In this paper, we calculated the current-current correlation with respect to the CME at two-loop level within the NJL model to check if there is higher order corrections to the CME current. Someone may argue that the CME coefficient is protected by anomaly so that  it is  non-renormalized. 
	This argument may come from the fact when one connects the general VVA triangle diagram, which is protected by the Adler-Bardeen theorem, with the CME current by expanding the current-current correlation in terms of $\mu_5$. However, we should emphasize that the triangle of CME is not exactly the general triangle but requires a vanishing momentum on the axial vertex. Since the VVA triangle is not IR safe on the axial vertex,  the current-current correlation might have the chance to get higher order 
	corrections. Although in the previous paper\cite{Hou2011Some}, the authors proved that the one-loop current-current correlation 
	vanished by the cancellation of the bare loop with its Pauli-Villars regularization, one may still doubt whether is was a general 
	case or just a coincidence. Actually, the answer to this question has been partly addressed in the section 4 of \cite{Hou2011Some}.
  Since we cannot place a confidence in the general relation 	between triangle anomaly and current-current correlation, an explicit calculation of higher order corrections is desired. That is  the reason why we do this two-loop calculation to the CME current. Fortunately, our result seems favour that the CME current is free from higher order corrections because the two-loop correction is still zero.
	
		The problem of higher order corrections to CME current is still far from solved since we only addressed the two-loop level within NJL model. A real QCD calculation is desired although it is rather complicated.
	Nevertheless, our calculation, as a toy model of QCD,  can give us confidence that one may finally find a way to proof that all higher order corrections vanish for	some reason.	

%гнгнгнгнгнгнгнгнгнгнгнгнгнгнгнгнгнгнгнгнгнгнгнгнгнгнгнгнгнгнгнгнгнгнгнгнгнгнгнгнгнгнгнгнгнгнгнгнгнгнгнгнгнгнгнгнгнгнгнгнгнгнгнгнгнгнгнгнгн	
\section*{Acknowledgements}
We thank Bo Feng, Hai-cang Ren and Defu Hou for their helpful
discussions and suggestions. This work is supported by
the National Natural Science Foundation of China under
Grant No. 11405074.

	\appendix
	
	\section{The Matsubara summation}
		  
		  In this appendix, the sum over the Matsubara energy $p_0$ or $k_0$ in the section III will be 
		  illustrated by an alternative method and an example is provided. 
		  
		  The summation of Matsubara energy $i\omega_n=(2n+1)\pi iT$ corresponding to the fermion can be replaced by a contour integral along the  
		  imaginary plane
		  \begin{equation}
			  M=T\sum\limits_{i\omega_n} D(i\omega_n)=\oint \frac{dz}{2\pi i} D(z) f(z)	
		  \end{equation}
		   with the Fermi distribution function
		  \begin{equation}
		 	  	f(z)=\frac{1}{e^{\beta z}+1}
		  \end{equation}
		  where the contour integral takes all poles that produced by the Fermi distribution function which equivalent to the summation. 
		  Deforming the contour to enclose singularities of $D(z)$, the summation can be completed by summing up residues of 
		  $D(z)f(z)$ over singularities of $D(z)$ that 		  
		  \begin{equation}
			  M=\sum\limits_{z_i} \textnormal{Res}D(z_i) f(z_i).
		  \end{equation}				  	  
		  
		  As an example, let us consider an  expression including singularities which reads 		  
		  \begin{equation}
		  	I_{\mu\nu}(P)={H_{\mu\nu}(p_0)\over {p_0^2-p^2}}
		  	\label{eq:eq5.3}
		  \end{equation}
		  where $H_{\mu\nu}$ is an arbitrary function. The sum over Matsubara energy $i\omega_n=(2n+1)\pi iT$ 
		  for fermion is provided by 
		  
		  \begin{equation}
		  	\begin{aligned}
		  		T\sum\limits_{p_0}I_{\mu\nu}(p_0,\textbf{p})=&T\sum\limits_{p_0}{H_{\mu\nu}(p_0)\over{(p_0+\textbf{p})}(p_0-\textbf{p})}\\
		  		=&\oint_C {dz\over {2\pi i}} {H_{\mu\nu}(z)\over{(z+\textbf{p})(z-\textbf{p})}} 
		  		{1\over{e^{\beta z}+1}}\\
		  		=&f(\textbf{-p}){H_{\mu\nu}(\textbf{-p})\over -2\textbf{p}}
		  		+f(\textbf{p}){H_{\mu\nu}(\textbf{p})\over 2\textbf{p}}.		  	
		  	\end{aligned}
		  	\label{eq:eq5.4}
		  \end{equation}

%\bibliography{CME}%
	

\end{document}